\documentclass[10pt]{amsart}
\usepackage{amsmath}
\usepackage{latexsym}
\usepackage{amsthm}
\usepackage{latexsym,amssymb}
\usepackage{amstext}
\usepackage{amsfonts}
\newcommand{\ignore}[1]{}
\newcommand{\R}{\mathbb R}

\newcommand{\rz}{\mathbb R}

\begin{document}
\newtheorem{thm}{Theorem}
\newtheorem{defn}[thm]{Definition}
\newtheorem{prop}[thm]{Proposition}
\newtheorem{ex}[thm]{Example}
\newtheorem{lem}[thm]{Lemma}
\newtheorem{cor}[thm]{Corollary}

\title{The Shapley Value of Phylogenetic Trees}

\author[Claus-Jochen Haake]{Claus-Jochen~Haake~$^1$}
\thanks{$^1$Institute of
  Mathematical Economics, Bielefeld University, PO Box 100131,
  33501 Bielefeld, Germany, {\tt chaake@wiwi.uni-bielefeld.de}}
\thanks{$^2$Department of Mathematics, Harvey Mudd
  College, Claremont, CA 91711, U.S.A., {\tt akashiwada@hmc.edu, su@math.hmc.edu}}
\author[Akemi Kashiwada]{Akemi~Kashiwada~$^{2,*}$} \thanks{$^*$Research
  partially
  supported by a Howard Hughes Medical Institute Undergraduate Science
  Education Program grant to Harvey Mudd
  College.}
\author[Francis E. Su]{Francis~Edward~Su~$^{2,**}$} \thanks{$^{**}$Research
  partially supported by NSF Grants DMS-0301129 and DMS-0701308.}

\begin{abstract}
  Every weighted tree corresponds naturally to a cooperative game that
  we call a {\em tree game};
  it assigns to each subset of leaves the
  sum of the weights of the minimal subtree spanned by those leaves.
  In the
  context of phylogenetic trees, the leaves are species and this
  assignment captures the {\em diversity} present in the coalition of
  species considered.  We consider the Shapley value of tree games and
  suggest a biological interpretation.
  We determine the linear transformation $\mathbf{M}$ that shows the
  dependence of the Shapley value on the edge weights of the tree, and
  we also compute a null space basis of $\mathbf{M}$.  Both depend on
  the {\em split counts} of the tree.  
  Finally, we characterize the Shapley value on tree games by four
  axioms, a counterpart to Shapley's original theorem on the larger class of
  cooperative games.  We
  also include a brief discussion of the core of tree games.

\end{abstract}


\dedicatory{revised, Aug 2007}

\maketitle


\section{Introduction}

The {\em Shapley value} is arguably the most important solution
concept for $n$-player cooperative games. Given a set of players $N$
of size $n=|N|$
in a cooperative game $v$, the Shapley value $\varphi(N,v)$ is the
unique imputation vector that satisfies four ``fairness'' criteria
(the {\em Shapley axioms}) that we shall discuss later.
In this paper we consider the game $v_\mathcal{T}$ induced by an
unrooted $n$-leaf tree $\mathcal{T}$ in which each edge is assigned
a positive number called an {\em edge weight}.  In this context, the
players are represented by the leaves of the tree and the value of
any coalition $S$ is the total weight of the subtree spanned by the
members of $S$.

In a more applied context, we consider games induced by a {\em
phylogenetic tree} in which players are species and the tree
represents a proposed evolutionary relationship among the species.
We suggest that a biological interpretation for the Shapley value is
a notion of the average marginal diversity that a species brings to
any group, and we study how the Shapley value depends on the edge
weights and topology of the tree.

One possible application of the Shapley value of a phylogenetic tree
is the economic theory of biodiversity preservation \cite{NePu02,
Weit92}.
In such contexts, quantifying the biological diversity of a species or a group of species is of great interest; many measures have been proposed (see, e.g., \cite{Fait92, MHCh05, PSOD05}) .
The {\em Noah's ark problem} \cite{Weit98, HaSt06} asks how to
prioritize species in a population if only some limited number can
be saved; we suggest that Shapley value provides a natural ranking
criterion 
as it is provides a measure of the contribution each species brings to the diversity of a group.

The literature applying game-theoretic solution concepts to an
analysis of trees appears to be limited.  One closely related
example is Kar \cite{Kar02}, who studies cost-sharing in a network
structure and characterizes the Shapley value of the minimum cost
spanning tree game of an arbitrary graph.
Also, \cite{myer77} as well as \cite{owen86} study values for games
that arise from a tree structure. However, these three works differ
from ours because there  each node of a graph is considered as a
player in the game, whereas we specifically study tree games and
allow only leaves as players.
Day and McMorris \cite{DaMc03} propose suitable axioms for a
consensus rule that will aggregate several phylogenetic trees into
one consensus tree; this differs from the thrust of our work, which
is to consider one tree and explore the interpretation and
properties of the Shapley value of the associated tree game.

In the next section we provide a biological interpretation for
the Shapley value of phylogenetic trees.  Then we discuss the
mathematics of calculating the Shapley value on tree games, starting
with some examples on small trees.  
We determine the linear transformation that shows how the Shapley
value depends on the edge weights of the tree, and compute a null
space basis that shows how to vary edge weights without changing the
Shapley value.  We also explain how these depend on the tree topology.
We conclude this paper by developing an analogue
of Shapley's theorem that characterizes the Shapley value on games by
four axioms.  We show that on the smaller class of tree games, the
Shapley value is characterized by those four axioms plus an additional
axiom.


\section{Phylogenetic Trees and the Shapley Value}

\subsection{Phylogenetic trees}

Evolutionary relationships between species are frequently
represented by a {\em phylogenetic tree}.  Evidence for such
relationships can come from a variety of sources, such as genomic
data or morphological comparisons, and much work has been done to
develop methods for constructing a phylogenetic tree from such data
(for surveys, see Felsenstein \cite{Fels04} and Semple-Steel
\cite{SeSt03}).

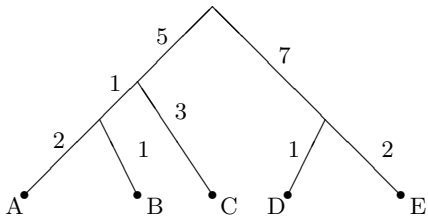
\begin{figure}[h]
\begin{center}
\setlength{\unitlength}{.5cm}
\begin{picture}(12,7)
\put(1,1){\line(1,1){5}}
\put(1,1){\circle*{.2}}
\put(4,1){\line(-1,2){1}}
\put(4,1){\circle*{.2}}
\put(6,1){\line(-2,3){2}}
\put(6,1){\circle*{.2}}
\put(8,1){\line(1,2){1}}
\put(8,1){\circle*{.2}}
\put(11,1){\circle*{.2}}
\put(11,1){\line(-1,1){5}}
\put(.5,.5){\small A}
\put(4.25,.5){\small B}
\put(6.2,.5){\small C}
\put(7.45,.5){\small D}
\put(11.25,.5){\small E}
\put(1.75,2.25){\small $2$}
\put(4,2){\small 1}
\put(3.25,3.75){\small 1}
\put(5,3){\small 3}
\put(4.5,5){\small 5}
\put(7.75,4.5){\small 7}
\put(8,2){\small 1}
\put(10.5,2){\small $2$}
\end{picture}
\caption{Example of a phylogenetic tree with species A-E with edge
  weights labeled.}\label{Fi:phylo}
\end{center}
\end{figure}

Phylogenetic trees are usually binary trees in which each internal
node represents a bifurcation in some characteristic and the leaves
are the species for which we have data.  Each edge has a weight that
represents some unit of distance between the nodes at its endpoints
(for instance, it could be the time between speciation events).
Figure \ref{Fi:phylo} gives a small example of what a (rooted)
phylogenetic tree could look like.  However, in this paper we shall
not be concerned with the location of the root of a tree, so all our
trees will be unrooted.

Formally, we shall think of a phylogenetic tree
${\mathcal T}$ as an unrooted tree with leaf set $N := \left\{
1,\ldots, n \right\}$ (representing the species in the population),
edge set $E$, and and an edge weight $\alpha_k$ for each edge $k$ in
$E$.


\subsection{The Shapley value}

In cooperative game theory, a {\em cooperative game} is a pair
$(N,v)$ consisting of a set of {\em players} $N=\{1,2,...,n\}$ and a
{\em characteristic function} $v$ that takes every subset of $N$
(called a {\em coalition}) to a real number (called the {\em worth}
of the coalition).  The subset consisting of all players is called
the {\em grand coalition}.  Formally, if $2^N$ is the set of all
subsets of $N$, then $v:2^N \rightarrow \rz$.
For instance, $N$ could be a set of companies and $v$ could describe
the profit that each coalition of companies could make if the
members of that coalition worked together.

One of the basic questions in cooperative game
theory is: if players work together to achieve some total worth (in
our example, profit), how should players then distribute their worth
(profit) among themselves?

As all (Pareto efficient) solution concepts from cooperative game
theory do, the {\em value} introduced by Shapley \cite{Shap53}
suggests a ``fair'' distribution of the total worth of the entire
set of players $N$ among the members of $N$.  Given a cooperative
game $(N,v)$, the Shapley value is a vector $\varphi = (\varphi_i)$
defined by the formula
\begin{equation}
\label{E:sv} \varphi_i(N,v) =
\frac{1}{n!}\sum_{\substack{S \subseteq
    N \\ i \in S}} (s-1)!(n-s)!(v(S)-v(S-i))
\end{equation}
where $s=|S|$ is the size of the coalition $S$ and $n=|N|$ is the
total number of players.

The formula above has a sensible interpretation that suggests a
rationale for the Shapley value to obtain a ``fair'' distribution.
For a player $i \in N$ and a coalition $S\subseteq N$ that contains
$i$, the quantity $v(S) - v(S - i)$ describes $i$'s marginal
contribution to the worth of $S$.  Then, if we choose an ordering of
the players (uniformly at random, in $n!$ ways) and if $i$ appears
as the $s$-th person in that order, then $i$'s marginal contribution
will be $v(S)-v(S-i)$ for each ordering in which the members of
$S-i$ appear before $i$ and the members of $N\setminus S$ appear
after $i$.  This may happen in $(s-1)! (n-s)!$ ways.  Hence the
combinatorial form of (\ref{E:sv}) reflects the Shapley value's
interpretation as the {\em expected marginal contribution} that $i$
makes.


\subsection{The Phylogenetic Tree Game}

Given a phylogenetic tree ${\mathcal T}$, we can define an
associated cooperative game $(N,v_{\mathcal T})$ that we call a {\em
phylogenetic tree game}.  Let $N$ be the set of leaves of the tree
(species).  For any subset $S\subseteq N$ of species, consider the
unique spanning subtree containing the members in $S$, and let
$v_{\mathcal T}(S)$ be the sum of the edge weights of that spanning
tree.  Thus for each set $S$ we may think of $v_{\mathcal T}(S)$ as a
measure of the {\em phylogenetic diversity} \cite{Fait92} within $S$.  This measure and its computational aspects have been studied much in recent years (see e.g., \cite{Stee05, MKvH06, PaGo05}).

Then the pair $(N,v_{\mathcal T})$ naturally forms a cooperative game.
Although species can hardly be compared with rationally acting
agents (as usually assumed in theory of cooperative games), we may
still ask for a meaningful re-interpretation of game-theoretic
solution concepts such as the Shapley value in the context of
phylogenetic trees.

Given a phylogenetic tree game $(N,v_{\mathcal T})$, equation
(\ref{E:sv}) suggests that the Shapley value of a given species may
be thought of as its {\em average marginal diversity}, i.e., the
average diversity the species can be expected to add to a group that
it joins.  So if $\varphi_i > \varphi_j$, then species $i$ can be
thought to contribute a greater diversity to a group than species
$j$ might.

\begin{ex}
\label{ex:five}

From direct calculations using (\ref{E:sv}), the five-leaf tree
${\mathcal{T}}$ in Figure \ref{Fi:Exfive} has Shapley value
 $$\varphi= (\varphi_A,\varphi_B,\varphi_C,\varphi_D,\varphi_E) = (5.28, 6.78, 4.2, 4.95, 2.78)$$ as we will show in Section \ref{Sec:fourfive}.

\begin{figure}[ht]
\begin{center}
\setlength{\unitlength}{.5cm}
\begin{picture}(10,6)

\put(1,1){\line(1,1){2}} \put(1,1){\circle*{.2}} \put(.25,.5){\small
B} \put(3,3){\circle*{.2}} \put(1,5){\circle*{.2}}
\put(1,5){\line(1,-1){2}} \put(.5,5.5){\small A}
\put(3,3){\line(1,0){4}} \put(7,3){\circle*{.2}}
\put(2.25,4.25){\small 3} \put(2.25,1.5){\small 5}
\put(4,2.5){\small 6} \put(6,2.5){\small 4} \put(5,3){\line(0,1){2}}
\put(5,3){\circle*{.2}} \put(5,5){\circle*{.2}}
\put(4.75,5.5){\small E} \put(5.25,4){\small 1}
\put(7,3){\line(1,1){2}} \put(7,3){\line(1,-1){2}}
\put(9,5){\circle*{.2}} \put(9,1){\circle*{.2}}
\put(7.75,4.25){\small 2} \put(7.75,1.5){\small 3}
\put(9.5,.5){\small D} \put(9.5,5.5){\small C}
\end{picture}
\caption{A five-leaf tree.}\label{Fi:Exfive}
\end{center}
\end{figure}
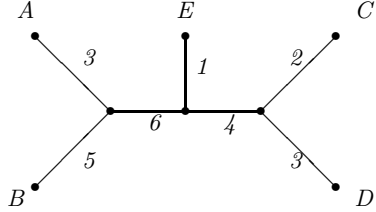

\end{ex}

\subsection{The Shapley Value Axioms}
\label{subsec:shapleyaxioms}

Besides the interpretation of the Shapley value as an average
expected marginal contribution, there is an axiomatization of the
Shapley value (see \cite{Shap53}) that uniquely characterizes it by
a set of (desirable) properties.  We review the axioms presented by
Shapley and discuss their plausibility in the present setting as
properties of phylogenetic trees.  Let therefore ${\mathcal V} :=
\left\{ v:2^N \rightarrow \rz \,|\,
  v(\emptyset) = 0  \right\}$ be the set of all cooperative games with $n$ players.

\begin{enumerate}

\item
({\em Pareto Efficiency Axiom})   The Shapley value is Pareto
efficient, i.e., $\sum_{i \in N} \varphi_i(N,v) = v(N)$ for all $v
\in \mathcal{V}$.

This axiom just states that the total diversity present within a
phylogenetic tree will be distributed and ascribed to the species
within it.  This is a reasonable axiom, given that the purpose of a
solution concept for a cooperative game is to distribute the worth
of the grand coalition among its members.  In this context, the
natural interpretation is that the Shapley value answers the
question of how much a specific species is responsible for the total
diversity, or, put another way, what is its {\em share} of
$v_{\mathcal T}(N)$.

\item ({\em Symmetry Axiom}) For any permutation of players $\pi:N \rightarrow N$
  the Shapley value satisfies $\varphi(\pi v) = \pi \varphi(v)$, where
  $\pi v$ is the permuted game given by $\pi v(S) := v(\pi^{-1}(S))$
  for all $S\subseteq N$ and $\pi \varphi(v)$ is the permuted solution
  vector, i.e., $(\pi\varphi(v))_i := \varphi_{\pi^{-1}(i)}(v)$.

The symmetry axiom states that a player's allocation should not be
  based on her name.  Another consequence of the symmetry axiom is
  if exchanging two players causes no difference in the worth
  that each adds to any coalition, then they should have the same Shapley value.
Biologically speaking, if two species play the same role within a
  tree then they should be ascribed the same responsibility for
  diversity, which seems to be a plausible requirement.

\item ({\em Dummy Axiom}) A dummy player is one that does not add worth to the
  value of any coalition.  This axiom says that dummy players should
  have a Shapley value of zero.

  This axiom is vacuously satisfied in the case of a phylogenetic
  tree game because there are no dummy species.  To see this, note
  that every species $i$ adds worth to the coalition that consists
  of a single species $j \neq i$, because the weight of the subtree
  containing $i$ and $j$ is the sum of the edge weights between $i$ and $j$
  and is therefore non-zero, but the weight of the subtree
  consisting of the singleton $j$ is zero.
(Even though there are no dummy species, this is still a reasonable
axiom here, since any species that does not diversify any coalition
should get value zero.)\footnote{In
Section~\ref{sec:char-shapl-value} we will replace the dummy axiom
by a different one to characterize the Shapley value on the class of
games that actually come from trees.}

\item ({\em Additivity Axiom}) Given two games $(N,v)$ and $(N,w)$ in ${\mathcal V}$
  with the same set
  of players $N$, define the {\em sum game} $(N,v+w)$ with characteristic
  function $(v+w)(S) = v(S) + w(S)$ for every coalition $S$.
  This axiom stipulates that the Shapley value of the sum game should be the
  sum of the Shapley values of the individual games:
  $\varphi(N,v+w) = \varphi(N,v) + \varphi(N,w)$.

  As an example, suppose we are given nucleotide sequences for a set of species $N$, and
  each sequence has length 200.  For each pair of species $i,j$ consider the (rather crude) measure
  of distance $d(i,j)$ to be the number of positions in which the sequences differ.
  The pairwise distance data can be used to construct a tree (using any standard
  method) and consequently, a tree game.  Thus the first 100 positions of the sequences can be used to
  construct a tree game $(N,v_1)$, and the second 100 positions a tree game $(N,v_2)$.  Then the
  Shapley value of the sum game $(N,v_1+v_2)$ is the sum of the Shapley values for each
  game.  This seems plausible in this context, since if the pairwise distances $d(i,j)$
  from both sets of 100 positions actually arise from a tree metrics on the same topological
  tree,
  then the sum game will arise from the tree reconstructed from all 200 positions.

\end{enumerate}


\section{Examples and Motivation: The Shapley Value for Small Trees}

As can be seen from (\ref{E:sv}), the Shapley value of a tree game
is a linear function of the edge weights of the tree.  We call that
linear transformation the {\em Shapley transformation}.  Before
deriving a general formula for this transformation in the subsequent
section, we study the Shapley transformation for games induced by
unrooted three-, four-, five- and six-leaf trees.

We will refer to the weights of edges incident to leaves as {\em
leaf weights} and other edge weights as {\em internal edge weights}.
Note that for an unrooted $n$-leaf tree, there are $n-2$ internal
nodes and $n-3$ internal edges in $E$.  In what follows, the
superscript $^T$ denotes the {\em transpose}.

\begin{defn}\label{Def:matrixsv}

  Let $\mathcal{T}$ be an $n$-leaf tree with leaves $N = \{1, \ldots,
  n\}$, associated leaf weights $\alpha_1, \ldots, \alpha_n$ and
  internal edges $I_1, \ldots, I_{n-3}$ with associated internal edge weights
  $\alpha_{I_1}, \ldots, \alpha_{I_{n-3}}$.
  Let $\vec{E}$ be a vector consisting of the edge weights in this order:
  $(\alpha_1, ..., \alpha_n, \alpha_{I_1}, ..., \alpha_{I_{n-3}})^T$.
  Define
  $\mathbf{M}=\mathbf{M}(N,v_{\mathcal T})$ to be the
  $n \times (2n-3)$ matrix that represents the Shapley transformation, so that
  the Shapley value of the game $v_{\mathcal{T}}$ is
$$\varphi(N,v_{\mathcal T}) = (\varphi_1, \varphi_2, \ldots, \varphi_n)^T = \mathbf{M} \vec{E}$$
where $\varphi_i$ is the Shapley value associated with leaf $i$.
Note that $\mathbf{M}$ depends on the topology of the $n$-leaf tree.

\end{defn}

Later in Theorem \ref{Thm:matrix} we determine a formula for
$\mathbf{M}[i,k]$, which is the coefficient of edge weight $k$ in
the calculation of the Shapley value of $i$.  But first, we give a
few examples.


\subsection{Three-Leaf Trees}\label{Sec:three}

\begin{figure}[ht]
\begin{center}
\setlength{\unitlength}{.5cm}
\begin{picture}(7,6)
\put(1,1){\line(1,1){2}}
\put(1,1){\circle*{.2}}
\put(.25,.5){\small B}
\put(3,3){\circle*{.2}}
\put(1,5){\circle*{.2}}
\put(1,5){\line(1,-1){2}}
\put(.25,4.95){\small A}
\put(3,3){\line(1,0){3}}
\put(6,3){\circle*{.2}}
\put(6.45,2.75){\small C}
\put(2.25,4.25){\small $\alpha$}
\put(2.25,1.5){\small $\beta$}
\put(4.25,3.25){\small $\gamma$}
\end{picture}
\caption{The topology of an unrooted three-leaf tree ${\mathcal{T}}$
  where the players are A, B, and C with corresponding leaf weights
  $\alpha$, $\beta$, and $\gamma$.}\label{Fi:three}
\end{center}
\end{figure}
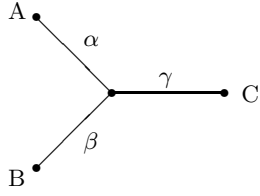

Topologically, there is only one unrooted three-leaf tree
${\mathcal{T}}$.  Let the leaves represent players A, B, and C with
corresponding leaf weights $\alpha$, $\beta$, and $\gamma$ as seen
in Figure \ref{Fi:three}.

The characteristic function $v_{\mathcal{T}}$ for this game is
$$v_{\mathcal{T}}(A) = v_{\mathcal{T}}(B) = v_{\mathcal{T}}(C) = 0,$$
$$v_{\mathcal{T}}(AB) = \alpha + \beta,\ \ v_{\mathcal{T}}(AC) =
\alpha + \gamma,\ \ v_{\mathcal{T}}(BC) = \beta + \gamma,$$
$$v_{\mathcal{T}}(ABC) = \alpha + \beta + \gamma.$$

Using Definition~\ref{Def:matrixsv}, we can calculate the Shapley
value by $\varphi = (\varphi_A, \varphi_B, \varphi_C) =
\mathbf{M}\vec{\ell}$ where $\vec{\ell}$ is the vector of leaf weights
$( \alpha, \beta, \gamma)^T$ and
$$\mathbf{M} = \frac{1}{6}
\left[
\begin{array}{ccc}
4 & 1 & 1 \\
1 & 4 & 1 \\
1 & 1 & 4
\end{array} \right].$$

It is apparent that we can solve for $\alpha$, $\beta$, and $\gamma$
in terms of $\varphi$ by inverting $\mathbf{M}$:
$$\vec{\ell} = \frac{1}{3}
\left[
\begin{array}{ccc}
5 & -1 & -1 \\
-1 & 5 & -1 \\
-1 & -1 & 5
\end{array}
\right]
\left(
\begin{array}{c}
    \varphi_A \\ \varphi_B \\ \varphi_C \end{array}\right).$$
This means the Shapley value of a 3-leaf tree uniquely determines
the tree representing the game.


\subsection{Four- and Five-Leaf Trees}\label{Sec:fourfive}

\begin{figure}[ht]
\begin{center}
\setlength{\unitlength}{.5cm}
\begin{picture}(21,6)
\put(1,1){\line(1,1){2}}
\put(1,1){\circle*{.2}}
\put(.25,.5){\small B}
\put(3,3){\circle*{.2}}
\put(1,5){\circle*{.2}}
\put(1,5){\line(1,-1){2}}
\put(.5,5.5){\small A}
\put(3,3){\line(1,0){3}}
\put(6,3){\circle*{.2}}
\put(2.25,4.25){\small $\alpha$}
\put(2.25,1.5){\small $\beta$}
\put(4.25,3.25){\small $\mu$}
\put(6,3){\line(1,1){2}}
\put(6,3){\line(1,-1){2}}
\put(8,5){\circle*{.2}}
\put(8,1){\circle*{.2}}
\put(6.75,4.25){\small $\gamma$}
\put(6.75,1.5){\small $\delta$}
\put(8.5,.5){\small D}
\put(8.5,5.5){\small C}

\put(12,1){\line(1,1){2}}
\put(12,1){\circle*{.2}}
\put(11.25,.5){\small B}
\put(14,3){\circle*{.2}}
\put(12,5){\circle*{.2}}
\put(12,5){\line(1,-1){2}}
\put(11.5,5.5){\small A}
\put(14,3){\line(1,0){4}}
\put(18,3){\circle*{.2}}
\put(13.25,4.25){\small $\alpha$}
\put(13.25,1.5){\small $\beta$}
\put(15,2.5){\small $\mu$}
\put(17,2.5){\small $\rho$}
\put(16,3){\line(0,1){2}}
\put(16,3){\circle*{.2}}
\put(16,5){\circle*{.2}}
\put(15.75,5.5){\small E}
\put(16.25,4){\small $\epsilon$}
\put(18,3){\line(1,1){2}}
\put(18,3){\line(1,-1){2}}
\put(20,5){\circle*{.2}}
\put(20,1){\circle*{.2}}
\put(18.75,4.25){\small $\gamma$}
\put(18.75,1.5){\small $\delta$}
\put(20.5,.5){\small D}
\put(20.5,5.5){\small C}
\end{picture}
\caption{(\emph{left})The topology for an unrooted four-leaf tree
  where the players are A, B, C, and D.  (\emph{right}) The unrooted
  five-leaf tree with players A, B, C, D, and E.}\label{Fi:four&five}
\end{center}
\end{figure}
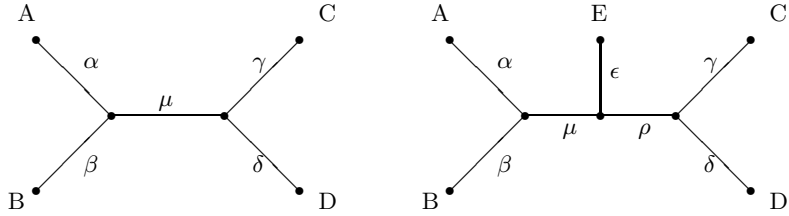

Similarly, we can calculate the Shapley value for each player in the
four- and five-leaf cases.  There is a unique tree topology for each
case, as shown in Figure~\ref{Fi:four&five}.

The Shapley value for the general four-leaf tree game is
$$\frac{1}{24} \left[\begin{array}{ccccc}
18 & 2 & 2 & 2 & 6 \\
2 & 18 & 2 & 2 & 6 \\
2 & 2 & 18 & 2 & 6 \\
2 & 2 & 2 & 18 & 6
\end{array} \right] \left( \begin{array}{c} \alpha \\ \beta \\ \gamma
  \\ \delta \\ \mu \end{array}\right).$$

Similarly for the five-leaf tree game, the Shapley value is
$$\frac{1}{120} \left[ \begin{array}{ccccccc}
96 & 6 & 6 & 6 & 6 & 36 & 16 \\
6 & 96 & 6 & 6 & 6 & 36 & 16 \\
6 & 6 & 96 & 6 & 6 & 16 & 36 \\
6 & 6 & 6 & 96 & 6 & 16 & 36 \\
6 & 6 & 6 & 6 & 96 & 16 & 16
\end{array}\right] \left(\begin{array}{c} \alpha \\ \beta \\ \gamma \\
  \delta \\ \epsilon \\ \mu \\ \rho \end{array} \right).$$
  This formula produces the calculation in Example \ref{ex:five}.

It is apparent from the fact that there are more variables (edge
weights) than equations that there is not a unique set of (possibly
negative) edge weights for a given Shapley value. That is, there is
not a unique tree corresponding to a given Shapley value.  The null
space of $\mathbf{M}$ will therefore help us determine which
weighted trees have the same Shapley value.  A basis for the null
space of $\mathbf{M}$ for the four-leaf tree is
$$\left\{ \left( \begin{array}{c} -1/4 \\ -1/4 \\ -1/4 \\ -1/4 \\ 1
    \end{array} \right) \right\}$$
This means that given a tree ${\mathcal{T}}$, we can produce other
trees with the same Shapley value by reducing the leaf weights by
$1/4$ for each unit increase in the internal edge weight.

Similarly, a null space basis for the five-leaf tree is
$$\left\{ \left( \begin{array}{c} -1/3 \\ -1/3 \\ -1/9 \\ -1/9 \\ -1/9
      \\ 1 \\ 0 \end{array} \right),\ \ \ \ \left( \begin{array}{c}
      -1/9 \\ -1/9 \\ -1/3 \\ -1/3 \\ -1/9 \\ 0 \\ 1 \end{array}
  \right) \right\}.$$

\begin{figure}[ht]
\begin{center}
\setlength{\unitlength}{.5cm}
\begin{picture}(10,6)

\put(1,1){\line(1,1){2}} \put(1,1){\circle*{.2}} \put(.25,.5){\small
$B$} \put(3,3){\circle*{.2}} \put(1,5){\circle*{.2}}
\put(1,5){\line(1,-1){2}} \put(.5,5.5){\small $A$}
\put(3,3){\line(1,0){4}} \put(7,3){\circle*{.2}}
\put(2.25,4.25){\small 4} \put(2.25,1.5){\small 6}
\put(4,2.5){\small 3} \put(6,2.5){\small 4} \put(5,3){\line(0,1){2}}
\put(5,3){\circle*{.2}} \put(5,5){\circle*{.2}}
\put(4.75,5.5){\small $E$} \put(5.25,4){\small $1 \frac{1}{3}$}
\put(7,3){\line(1,1){2}} \put(7,3){\line(1,-1){2}}
\put(9,5){\circle*{.2}} \put(9,1){\circle*{.2}}
\put(7.25,4.25){\small $2 \frac{1}{3}$} \put(7.25,1.5){\small $3
\frac{1}{3}$} \put(9.5,.5){\small $D$} \put(9.5,5.5){\small $C$}
\end{picture}
\caption{A five-leaf tree with same Shapley value as Figure
\ref{Fi:Exfive}.}
\label{Fi:Exfive-alt}
\end{center}
\end{figure}
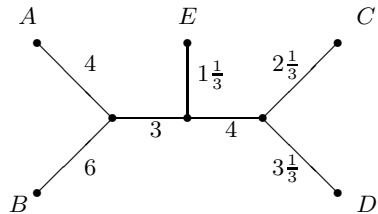

For example, the first basis element (multiplied by -3) shows us
that the tree in Figure \ref{Fi:Exfive-alt} 
has the same Shapley value as the tree in Figure \ref{Fi:Exfive}.

\subsection{Six-Leaf Trees}\label{Sec:six}

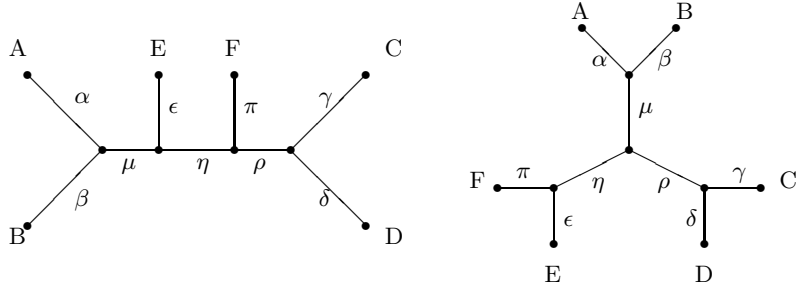
\begin{figure}[ht]
\begin{center}
\setlength{\unitlength}{.5cm}
\begin{picture}(22,8)
\put(1,2){\line(1,1){2}}
\put(1,2){\circle*{.2}}
\put(.5,1.5){\small B}
\put(3,4){\circle*{.2}}
\put(1,6){\circle*{.2}}
\put(1,6){\line(1,-1){2}}
\put(.5,6.5){\small A}
\put(3,4){\line(1,0){5}}
\put(8,4){\circle*{.2}}
\put(2.25,5.25){\small $\alpha$}
\put(2.25,2.5){\small $\beta$}
\put(3.5,3.5){\small $\mu$}
\put(7,3.5){\small $\rho$}
\put(5.5,3.5){\small $\eta$}
\put(4.5,4){\line(0,1){2}}
\put(4.5,4){\circle*{.2}}
\put(4.5,6){\circle*{.2}}
\put(4.25,6.5){\small E}
\put(4.75,5){\small $\epsilon$}
\put(6.5,4){\line(0,1){2}}
\put(6.5,4){\circle*{.2}}
\put(6.5,6){\circle*{.2}}
\put(6.25,6.5){\small F}
\put(6.75,5){\small $\pi$}
\put(8,4){\line(1,1){2}}
\put(8,4){\line(1,-1){2}}
\put(10,6){\circle*{.2}}
\put(10,2){\circle*{.2}}
\put(8.75,5.25){\small $\gamma$}
\put(8.75,2.5){\small $\delta$}
\put(10.5,1.5){\small D}
\put(10.5,6.5){\small C}

\put(13.5,3){\line(1,0){1.5}}
\put(13.5,3){\circle*{.2}}
\put(15,1.5){\line(0,1){1.5}}
\put(15,1.5){\circle*{.2}}
\put(15,3){\circle*{.2}}
\put(12.75,3){\small F}
\put(14.75,.5){\small E}
\put(14,3.25){\small $\pi$}
\put(15.25,2){\small $\epsilon$}
\put(15,3){\line(2,1){2}}
\put(17,4){\circle*{.2}}
\put(17,4){\line(0,1){2}}
\put(17,6){\circle*{.2}}
\put(17,6){\line(-1,1){1.25}}
\put(17,6){\line(1,1){1.25}}
\put(15.75,7.25){\circle*{.2}}
\put(18.25,7.25){\circle*{.2}}
\put(17,4){\line(2,-1){2}}
\put(19,3){\line(1,0){1.5}}
\put(19,3){\line(0,-1){1.5}}
\put(19,3){\circle*{.2}}
\put(20.5,3){\circle*{.2}}
\put(19,1.5){\circle*{.2}}
\put(16,3){\small $\eta$}
\put(17.75,3){\small $\rho$}
\put(18.5,2){\small $\delta$}
\put(18.75,.5){\small D}
\put(19.75,3.25){\small $\gamma$}
\put(21,3){\small C}
\put(17.25,5){\small $\mu$}
\put(15.45,7.5){\small A}
\put(18.25,7.5){\small B}
\put(16,6.25){\small $\alpha$}
\put(17.75,6.25){\small $\beta$}
\end{picture}
\caption{(\emph{left})The first topology for an unrooted six-leaf tree
  $\mathcal{T}$ where the players are A, B, C, D, E and F.
  (\emph{right}) The second unrooted six-leaf tree
  $\mathcal{T}'$.}\label{Fi:six}
\end{center}
\end{figure}

For our last direct calculation, let us consider the games represented
by six-leaf trees.  In this case there are two topologies for unrooted
trees with six leaves (see figure \ref{Fi:six}).

The Shapley value for the first and second six-leaf trees are,
respectively,
$$\varphi(N,v_{\mathcal{T}})=\frac{1}{720} \left[ \begin{array}{ccccccccc}
600 & 24 & 24 & 24 & 24 & 24 & 240 & 60 & 120 \\
24 & 600 & 24 & 24 & 24 & 24 & 240 & 60 & 120 \\
24 & 24 & 600 & 24 & 24 & 24 & 60 & 240 & 120 \\
24 & 24 & 24 & 600 & 24 & 24 & 60 & 240 & 120 \\
24 & 24 & 24 & 24 & 600 & 24 & 60 & 60 & 120 \\
24 & 24 & 24 & 24 & 24 & 600 & 60 & 60 & 120
\end{array}\right]\left(\begin{array}{c} \alpha \\ \beta \\ \gamma \\
  \delta \\ \epsilon \\ \pi \\ \mu \\ \rho \\ \eta
\end{array}\right),$$
$$\varphi(N,v_{\mathcal{T}'})=\frac{1}{720} \left[ \begin{array}{ccccccccc}
600 & 24 & 24 & 24 & 24 & 24 & 240 & 60 & 60 \\
24 & 600 & 24 & 24 & 24 & 24 & 240 & 60 & 60 \\
24 & 24 & 600 & 24 & 24 & 24 & 60 & 240 & 60 \\
24 & 24 & 24 & 600 & 24 & 24 & 60 & 240 & 60 \\
24 & 24 & 24 & 24 & 600 & 24 & 60 & 60 & 240 \\
24 & 24 & 24 & 24 & 24 & 600 & 60 & 60 & 240
\end{array}\right]\left(\begin{array}{c} \alpha \\ \beta \\ \gamma \\
  \delta \\ \epsilon \\ \pi \\ \mu \\ \rho \\ \eta
\end{array}\right).$$

As with the four and five leaf cases, both topologies of the six
leaf tree allow for many trees to possess the same Shapley value.
The basis for the null space of the first six-leaf tree is
$$\left\{ \left( \begin{array}{c} -3/8 \\ -3/8 \\ -1/16 \\ -1/16 \\
      -1/16 \\ -1/16 \\ 1 \\ 0 \\ 0 \end{array} \right), \left(
    \begin{array}{c} -1/16 \\ -1/16 \\ -3/8 \\ -3/8 \\ -1/16 \\ -1/16
      \\ 0 \\ 1 \\ 0 \end{array} \right), \left( \begin{array}{c} -1/6
      \\ -1/6 \\ -1/6 \\ -1/6 \\ -1/6 \\ -1/6 \\ 0 \\ 0 \\ 1
    \end{array} \right) \right\}$$
and for the second six-leaf tree is
$$\left\{ \left( \begin{array}{c} -3/8 \\ -3/8 \\ -1/16 \\ -1/16 \\
      -1/16 \\ -1/16 \\ 1 \\ 0 \\ 0 \end{array} \right), \left(
    \begin{array}{c} -1/16 \\ -1/16 \\ -3/8 \\ -3/8 \\ -1/16 \\ -1/16
      \\ 0 \\ 1 \\ 0 \end{array} \right), \left( \begin{array}{c}
      -1/16 \\ -1/16 \\ -1/16 \\ -1/16 \\ -3/8 \\ -3/8 \\ 0 \\ 0 \\ 1
    \end{array} \right) \right\}.$$


\subsection{Notes on Relationship between Trees and Shapley Values}
\label{SS:treenotes}

From these examples, we make a few observations.
\begin{enumerate}
\item Any Shapley value $n$-vector can be realized by adjusting the
  edge weights of an $n$-leaf tree.
  This may involve positive as well as nonpositive edge weights.
  However, the positive hull of the column vectors of the matrix ${\bf
    M}$ can be realized as the Shapley value of trees with nonnegative
  edge weights.
\item When $n \geq 4$, there is not a unique $n$-leaf tree
  corresponding to a given Shapley value because the null space is
  nontrivial.
\item The null spaces for the two six-leaf trees are different (since
  there is exactly one basis for each null space whose projection to
  the last 3 coordinates are the standard unit vectors in $\R^3$, and
  these bases are different for the given null spaces).  Moreover,
  one may check that there is no permutation of the coordinates of one
  null space that will identify it with the other null space (we say
  such spaces are not {\em permutation equivalent}).  As we shall see
  in Section \ref{SS:null}, if the null spaces are not permutation
  equivalent, then the two trees must not be isomorphic (Theorem
  \ref{Thm:nonisomorphic}).
\item Under close inspection, one notices a
  relationship between the numbers of leaves on each side of an internal
  edge (the {\em split counts}) and quantities such as the entries of the Shapley transformation
  matrix and the null space basis vectors.  We exhibit their
  explicit dependence in the following sections.
\end{enumerate}


\section{The Shapley Transformation}

We first show the contribution of each edge weight to the Shapley
value; these are the entries of the matrix $\mathbf{M}$ representing
the Shapley transformation.  The following theorem gives us a quick
way of finding the $(i,k)$th entry of $\mathbf{M}$.  Before we state
and prove the theorem, we need a definition that will be
instrumental throughout the rest of this paper.

\begin{defn}
  Let $\mathcal{T}$ be an $n$-leaf tree with leaves $N$ and edges $E$.
  For $i \in N$ and $k \in E$, the removal of edge $k$ splits
  $\mathcal{T}$ into two subtrees.
  Let $\mathcal{C}(i,k)$ denote the set of leaves in the subtree that {\em contains} $i$
  (the ``containing'' subtree) and let $\mathcal{F}(i,k)$ denote the
  set of leaves in the other subtree that is ``far'' from $i$.
  We then denote the number of leaves of
  $\mathcal{C}(i,k)$ and $\mathcal{F}(i,k)$ as $c(i,k)$ and $f(i,k)$,
  respectively.
\end{defn}

If it is obvious what leaf $i$ and edge $k$ we are referring to, we
will simply write $c,f$ instead of $c(i,k), f(i,k)$. Note that $n =
c+f$.  We call $c,f$ the {\em split counts} associated with leaf $i$
and edge $k$.  As we shall see, the split counts will arise
frequently in our results on the Shapley transformation.

\begin{thm}
\label{Thm:matrix}

  Let ${\mathcal{T}}$ be an $n$-leaf tree.
  The $(i,k)$th entry of the Shapley transformation
  matrix $\mathbf{M}$ is given by
  $$\mathbf{M}[i,k] = \frac{f(i,k)}{n\ c(i,k)}.$$

\end{thm}

\begin{proof}
  Fix leaf $i$.  To
  count the number of times a given edge weight contributes to $i$'s
  Shapley value, we need to know how many times it is in the marginal
  contribution of $i$ for coalitions of size $s$.  Edge weight
  $\alpha_k$ will be part of $i$'s marginal contribution if the other
  $s-1$ members of the coalition are from the far side of the
  edge from $i$.  So
  $$\mathbf{M}[i,k] = \frac{1}{n!} \sum_{s=2}^n (s-1)!(n-s)!{f(i,k) \choose
    s-1} = \frac{1}{n!} \sum_{s=2}^n \frac{(n-s)! f(i,k)!}{(f(i,k)-s+1)!}.$$
  Using the fact $f=n-c$, the above expression can be rewritten:
  $$\frac{1}{n!}\sum_{s=2}^n (n-c)!(c-1)!{n-s \choose c-1} =
  \frac{(n-c)!(c-1)!}{n!}\sum_{j=1}^{n-1} {j-1 \choose c-1}.$$
  We use the identity
  $$\sum_{j=1}^{n}{j-1 \choose c-1} = {n \choose c} = {n-1 \choose
    c-1}\frac{f}{c} + {n-1 \choose c-1}$$
  to obtain
  $$\mathbf{M}[i,k] = \frac{(n-c)!(c-1)!}{n!}{n-1 \choose c-1}\frac{f}{c} =
  \frac{f}{n c}.$$

\end{proof}

This result is particularly nice because it shows how the Shapley
value's dependence on any edge weight simply hinges on the number of
leaves on either side of that edge. Consider the following example.

\begin{ex}

  Using Theorem \ref{Thm:matrix} we will calculate the coefficient of
  $\mu$ in player A's Shapley value for a five-leaf tree.  Let the edge with
  edge weight $\mu$ be $I_1$.  There are three leaves in
  $\mathcal{F}(A,I_1)$ and two leaves in $\mathcal{C}(A,I_1)$.
  Thus,
$$\mathbf{M}[1,6] = \frac{3}{5 \cdot 2}$$
which is the same as the $(A,\mu)$ entry $36/120$ in the Shapley
transformation of the five-leaf tree given in section
\ref{Sec:fourfive}.

\end{ex}


\section{The Null Space of the Shapley Transformation}\label{SS:null}

Now we will also use Theorem~\ref{Thm:matrix} to understand the dependence of
the null space of the Shapley transformation on the split counts, as
suggested in Section \ref{SS:treenotes}.

The following theorem exhibits a null space basis of $\mathbf{M}$
in terms of the split counts.

\begin{thm}\label{Thm:null}

  Let $\mathcal{T}$ be an $n$-leaf tree with leaves $N = \{1, \ldots, n\}$ and internal edges $I_1, \ldots,
  I_{n-3}$.  The dimension of the null space of $\mathbf{M}=\mathbf{M}(N,v_{\mathcal
  T})$ is $n-3$.  A basis for the null space is the collection of vectors $\{ w_{I_k} \}$
  in $\mathbb{R}^{2n-3}$, one for each internal edge $I_k$:
\begin{equation}\label{E:null}
(w_{I_k})_i = \begin{cases}
-\frac{f(i,k)-1}{(n-2)c(i,k)} & \text{ if } 1 \leq i \leq n  \\
1 & \text{ if } i = n+k \\
0 & \text{ otherwise}
\end{cases}
\end{equation}
for all $k \in \{1, \ldots, n-3\}$ and entries $i \in \{1, \ldots,
2n-3\}$, where the first $n$ entries correspond to leaves and the
last $n-3$ entries correspond to internal edges.
\end{thm}

Before proving the theorem, we give an example.

\begin{ex}

  Consider the five-leaf tree in Figure \ref{Fi:four&five}.
  Let $I_1, I_2$ be the internal edges with weight $\mu,\rho$, respectively.
  We use Theorem
  \ref{Thm:null} to determine the null space vector $w_{I_1}$.
  The $5+1=6$-th entry of $w_{I_1}$ is 1 and all entries
  after that are 0.  To find the first five entries of the vector,
  consider the two subtrees obtained by removing $I_1$ from the
  tree, namely, the subtrees AB and CDE.  By (\ref{E:null}),
  the first two entries of the $w_{I_1}$
  corresponding to A and B will be
$$-\frac{3-1}{(5-2)2} = -\frac{1}{3}$$
and the next three entries corresponding to C, D, and E are
$$-\frac{2-1}{(5-2)3} = - \frac{1}{9}.$$
This agrees with the first null space basis vector we exhibited in
Section \ref{Sec:fourfive}.  (The other basis vector there is
$w_{I_2}$.)
\end{ex}

Now we prove Theorem \ref{Thm:null}.

\begin{proof}

  Let $\mathcal{T}$ be an $n$-leaf tree.  Consider the $i$th leaf.  If we
  let $\mathbf{M}$ be the matrix of Shapley value coefficients for $\mathcal{T}$
  then we want to show
\begin{equation}\label{E:dot}
\sum_{j=1}^{2n-3} \mathbf{M}[i,j] (w_{I_k})_j = 0.
\end{equation}

Fix $k \in \{1, \ldots, n-3\}$.  Note that by
Theorem~\ref{Thm:matrix}, for all leaves $j \neq i$,
$\mathbf{M}[i,j] = \frac{1}{n (n-1)}$ and $\mathbf{M}[i,i] =
\frac{n-1}{n}$.
The only other non-zero entry of $\mathbf{M}$ we need to consider is
the one associated with the $(n+k)$-th edge of the tree, i.e., the
internal edge $I_k$.  The split counts for $I_k$ are $c,f$ and so
$$\mathbf{M}[i,n+k] = \frac{f}{n \ c}.$$
Thus, showing (\ref{E:dot}) is the same as showing that
$$
  -f \cdot \frac{1}{n(n-1)} \cdot \frac{c-1}{(n-2)f}
  - (c-1) \cdot \frac{1}{n(n-1)} \cdot \frac{f-1}{(n-2)c}
  - \frac{n-1}{n} \cdot \frac{f-1}{(n-2)c}
  + \frac{f}{n \ c} = 0, \notag
$$
which can be checked by algebraic manipulation and the fact that
$n=f+c$.

\ignore{**********************
 But algebraic manipulations (and the fact that $n=f+c$) reveal
that
\begin{align}
& f \cdot \frac{1}{n(n-1)} \cdot \frac{c-1}{(n-2)f}
  + (c-1) \cdot \frac{1}{n(n-1)} \cdot \frac{f-1}{(n-2)c}
  + \frac{n-1}{n} \cdot \frac{f-1}{(n-2)c} \\
&= \left( \frac{(c-1)(n-1)}{n(n-1) (n-2)c} \right)
+\frac{(n-1)(f-1)}{n(n-2)c} \notag \\
&= \frac{f(n-2)}{n(n-2)c} \notag \\
&= \frac{f}{n c}, \notag
\end{align}
as desired. *************************}

Thus $w_{I_k}$ is in the null space of the Shapley transformation
$\mathbf{M}$. It is apparent that the null space has dimension $n-3$
and the $w_{I_k}$ are linearly independent. Therefore the $w_{I_k}$
form a basis of the null space of $\mathbf{M}$.
\end{proof}

We note that this basis $\{w_{I_k}\}$, $k=1,...,n-3$, is uniquely determined by fixing the last $n-3$ coordinates to be the standard basis vectors in $\R^{n-3}$.  For this reason we refer to this basis as the {\em standard} null space basis for $\mathbf{M}$.

An immediate corollary of Theorem \ref{Thm:null} is that the standard basis
of $Null(\mathbf{M})$ reveals which pairs of leaves form {\em cherries}. A pair
of leaves $(i,j)$ is called a {\em cherry} if they have a common
parent. This is the case if and only if the tree spanned by $i$ and
$j$ does not include an internal edge. Therefore, removing the
internal edge $k$ that contains the common parent will split the
tree into a 2-leaf subtree and an $(n-2)$-leaf subtree, and in such
a situation (as long as $n\geq 4$) we expect to find exactly two
entries in $w_{I_k}$ whose values $-(n-3)/2(n-2)$ correspond to the
two cherry leaves.
 \ignore{******************
\begin{cor} \label{thm:cherry}
Let ${\mathcal T}$ be an unrooted tree with leaves set $N$, $|N|\geq
4$, and edge set $E$. Let $w^k := w_{I_k} = (w^k_1,\ldots,
w^k_n,w^k_{n+1},\ldots,
  w^k_{2n-3})$ denote the basis vectors of the nullspace of the
  Shapley transformation as in (\ref{E:null}) .  Then a pair $(i,j)$ of leaves forms a cherry
  if and only if there exists $k$ such that
\begin{equation}\label{eq:cherries}
  w^{k}_i = w^{k}_j = -\frac{n-3}{2(n-2)}.
\end{equation}
\end{cor}
************************}
 The examples in Section \ref{Sec:fourfive}
and Section \ref{Sec:six} nicely illustrate this fact.

Call two trees {\em isomorphic} if there is a bijection between 
edges that takes one tree to the other and preserves the
topological structure of the tree.  Call two matrices {\em
  permutation-equivalent} if one can be obtained from the other by a
permutation of the rows and a permutation of the columns.  Call two
subspaces of $\R^n$ {\em permutation-equivalent} if one set can be
obtained from the other by some permutation of the coordinates.

Since the split counts of a tree only depend on the topology of the tree, Theorem \ref{Thm:matrix} shows that isomorphic trees will produce the same Shapley transformation matrix $\mathbf{M}$ up to a permutation of the rows (given by permuting the order of leaves that define the rows) and a permutation of the columns (given by a permuting the order of the edges that define the columns).  
The null space of $\mathbf{M}$ is not affected by permuting the rows of $\mathbf{M}$, but permuting the columns of $\mathbf{M}$ has the effect of permuting the coordinates of the null space of $\mathbf{M}$.  Therefore we summarize:

\begin{thm}
\label{Thm:nonisomorphic}
Isomorphic trees induce permutation-equivalent Shapley transformation matrices with permutation-equivalent null-spaces.  Hence, if for two trees $\mathcal{T}_1, \mathcal{T}_2$, their Shapley transformation matrices $\mathbf{M}_1,\mathbf{M}_2$ or their null spaces are not permutation-equivalent, then $\mathcal{T}_1, \mathcal{T}_2$ must not be isomorphic.
\end{thm}


\section{Characterization of the Shapley Value of Tree Games}
\label{sec:char-shapl-value}

The Shapley axioms presented in Section \ref{subsec:shapleyaxioms}
uniquely characterize the Shapley
value on the class of all $n$-person games. However, the class of
$n$-person games that are derived from a tree is smaller.
Thus while the Shapley axioms still hold for this smaller class, they
may no longer uniquely determine the Shapley value as a function on this
class.  In this section we will therefore strengthen the axioms so
that they once again uniquely characterize the Shapley value on the
class of $n$-person games derived from a tree.

By ${\mathcal V}^{N,E}$ we denote the class of games arising from some tree
with set of leaves $N$ and edge set $E$. For games in ${\mathcal
  V}^{N,E}$ we will allow positive as well as non-positive edge
weights. Thus, ${\mathcal V}^{N,E}$ is a linear space and we ask for its
dimension.  

For a fixed pair $(N,E)$ define games $v_k$ ($k\in
E)$ in the following way: $v_k$ corresponds to the tree in which
edge $k$ is weighted 1 and all other edges are weighted zero. We call
such a game a {\em basis game}. It is readily checked that the game
$v$ associated with the tree that exhibits edge weights
$\alpha_1,\ldots, \alpha_n,\alpha_{I_1},\ldots, \alpha_{I_{n-3}}$ is
the linear combination $v = \sum_{k\in E} \alpha_k v_k$. Moreover, the
family $(v_k)_{k\in E}$ is linearly independent. Therefore these games
form a basis of ${\mathcal V}^{N,E}$ and $\dim {\mathcal V}^{N,E} = 2n -
3$.

Note that in this context, the Shapley transformation $\mathbf{M}$, as
a $n\times (2n-3)$ matrix, can be viewed as a linear transformation
from ${\mathcal V}^{N,E}$ to $\R^n$.

To characterize this transformation, we ask what properties (axioms)
we might expect a "diversity measure'' $\psi:{\mathcal V}^{N,E} \rightarrow \R^n$ to satisfy.  Thus, given a tree game $v$, $\psi(v)$ is a vector in $\R^n$ which specifies for each of $n$ leaves (players) a number that measures, in some sense, their contribution to the diversity of a group.
For instance, for the basis game $v_k$, let us consider what a ``reasonable''
distribution $\psi(v_k) \in \rz^n$ might be.  We may interpret zero edge
weights on either side of the edge $k$ in the basis game $v_k$ as having two groups of
species, each one being homogeneous. So a natural property would be
that the degree of diversity that we assign to one group only
depends on the size of this group (and hence the size of the other
group) relative to the whole population. It seems plausible that a
given group on one side of an edge diversifies the population more
if there are more species on the other side of that edge.  Thus we
may assume that $\psi_i(v_k)$ is described by a function that is
increasing in the fraction $f(i,k)/n$. We formulate these
considerations as an additional axiom.

\medskip

{\bf Axiom (group proportionality on basis games):} For fixed $N$
and $E$, a mapping $\psi:{\mathcal V}^{N,E}$ is said to satisfy {\em
group proportionality on basis games}, if there is some constant $d
\in \rz$ such that $\psi$ satisfies $\sum_{j\in
  \mathcal{C}(i,k)}\psi_j(v_k) = d\,\frac{f(i,k)}{n}$ for all $i\in N, k\in
E$.

\medskip

Thus, with $\psi$ satisfying this axiom, a group's assigned
diversity linearly changes with the other group's fraction of the
whole population. Using the new axiom, we get a characterization
result for the Shapley value of games in $\mathcal{V}^{N,E}$, which
may be regarded as a counterpart to Shapley's original theorem
\cite{Shap53} characterizing the Shapley value on all games.

\begin{thm} \label{thm:characterization}
  For each pair $(N,E)$ (consisting of leaf set $N$ and edge set $E$)
  there is one and only one mapping $\psi:{\mathcal V}^{N,E} \rightarrow
  \rz^n$ that satisfies Pareto efficiency, symmetry, additivity and
  group proportionality. 
  This mapping coincides with the Shapley value $\varphi$  
  restricted to ${\mathcal V}^{N,E}$, i.e., $\psi(v_{\mathcal T}) = \varphi(N,v_{\mathcal T})$
  based on the phylogenetic diversity function $v_{\mathcal T}$.
\end{thm}

\begin{proof}
  It is immediately verified that the Shapley value satisfies all the
  axioms (for group proportionality, use Theorem \ref{Thm:matrix}).

  Now, let $(N,E)$ be fixed and $\psi$ satisfy the axioms. First, we
  take a basis game $v_k$ and determine $\psi$. By symmetry, we may
  conclude $\psi_i(v_k)=\psi_{j}(v_k)$ as long as $i,j$ are on the
  same side of edge $k$.  Hence
  \begin{equation} \label{sumclose}
  \sum_{j\in \mathcal{C}(i,k)} \psi_j(v_k) = c(i,k) \, \psi_i(v_k).
  \end{equation}
  Pareto
  efficiency and group proportionality imply $$v_k(N) = 1 = \sum_{j\in N} \psi_j(v_k) =
  \sum_{j\in \mathcal{C}(i,k)} \psi_j(v_k) + \sum_{j\in \mathcal{F}(i,k)} \psi_j(v_k) =
  d\,(\frac{f(i,k)}{n}+\frac{c(i,k)}{n}) =
  d.$$
   Hence $d=1$ and by group proportionality and (\ref{sumclose}),
   we obtain $\psi_i(v_k) = \frac{f(i,k)}{n \, c(i,k)}$ for
  any $i\in N$ and $k\in E$. Analogously, we get $\psi_i(\lambda v_k)
  = \lambda \, \psi_i(v_k)$ for $\lambda \in \rz$.
  Using additivity and
  Theorem~\ref{Thm:matrix}, $\psi$ coincides with the Shapley value on
  ${\mathcal V}^{N,E}$.
\end{proof}

We close this section with two remarks. First, note that any game
arising from a tree with nonnegative edge weights is representable as
a linear combination of basis games using nonnegative coefficients.
Hence, we may derive a version of
Theorem~\ref{thm:characterization} for classes of games that actually
arise from phylogenetic trees.

Second, Theorem~\ref{thm:characterization} provides further
justification for the use of the Shapley value to analyze phylogenetic
trees. If one wants to distribute the total diversity of a
population on its species and the distribution rule should satisfy the
above (reasonable) axioms, then the Shapley value is the only possible
choice. As symmetry, Pareto efficiency and additivity are rather
``obligatory'' requirements for a plausible rule, it is the
proportionality axiom that provides further insight in the rationale
behind the Shapley value. Of course, modification of the group
proportionality axiom eventually leads to a different distribution
rule based on a different rationale.


\section{The Core of Tree Games}\label{S:core}

In prior sections, we have explored the Shapley value as a solution
concept for tree games. However, another solution concept for
$n$-player cooperative games that is frequently studied is the {\em
  core} of a game, which is the set of all imputations $\vec{x} \in
\mathbb{R}^n$ such that for all coalitions $S \subseteq N$, $\sum_{i
  \in S} x_i \geq v(S)$ and $\sum_{i \in N} x_i = v(N)$. In this
section we examine the core of phylogenetic tree games.

We start with three- and four-leaf tree games for intuition.

\begin{ex}
  The characteristic function of the three-leaf tree game is given in
  Section \ref{Sec:three}, and yields the following system of
  inequalities for the core:
\begin{align}
  x_A+x_B+x_C &= \alpha + \beta + \gamma \notag \\
  x_A+x_B &\geq \alpha + \beta \notag \\
  x_A+x_C &\geq \alpha + \gamma \notag \\
  x_B+x_C &\geq \beta + \gamma \notag
\end{align}
Hence the core consists of the single element $\vec{\ell}$, the
vector of leaf weights $\left( \begin{array}{c} \alpha \\ \beta \\
    \gamma \end{array}\right)$.

\end{ex}

Thus the three-leaf tree has only one element in its core, namely
the vector of leaf weights.  Now we consider the four-leaf tree
game, which, unlike the three-leaf tree, has an internal edge.

\begin{ex}
The characteristic function of the
  four-leaf tree game in Figure \ref{Fi:four&five} yields the following system of
  inequalities for the core:
\begin{align}
  x_A+x_B+x_C+x_D &= \alpha + \beta + \mu + \gamma + \delta \notag \\
  x_A+x_C &\geq \alpha + \mu + \gamma \label{InE:AC} \\
  x_B+x_D &\geq \beta + \mu + \delta \label{InE:BD} \\
  \vdots \notag
\end{align}
 From (\ref{InE:AC}) and (\ref{InE:BD}) we see that
$$\alpha + \mu + \gamma \leq x_A + x_C \leq \alpha + \gamma.$$
So either $\mu = 0$ in which case we have a {\em degenerate} tree 
(internal edge weight zero) and the
core is $\vec{\ell}$, or the core has to be empty since the
inequality cannot be satisfied.
\end{ex}

These two examples illustrate the following theorem:
\begin{thm}
  Let ${\mathcal{T}}$ be an $n$-leaf game tree ${\mathcal{T}}$ where
  $n \geq 3$.  If the tree is degenerate (all internal edge weights
  are zero), then the core consists of the leaf weight vector
  $\vec{\ell}$.  Otherwise the core is empty.
\end{thm}

\begin{proof}
  Let ${\mathcal{T}}$ be an $n$-leaf tree with edge weights $\alpha_i$
  for $i \in \{1, \ldots, 2n-3\}$.  Every tree has at least two
  cherries, where a {\em cherry} is a set of two leaves with a common
  parent.  Label the two leaves on one cherry 1 and 2 and label the
  two leaves on the other cherry 3 and 4 each with corresponding leaf
  weights $\alpha_1$, $\alpha_2$, $\alpha_3$ and $\alpha_4$.  We know
  from the properties of the core that for the set of leaves $N$,
\begin{align}
\sum_{j \in N} x_j &= \sum_{i \in \{1, \ldots, 2n-3\}} \alpha_i
\label{E:total} \\
x_1 + x_3 &\geq \sum_{k \in P} \alpha_{k} \label{E:ACgeq} \\
\sum_{j \in N\setminus \{1,3\}} x_j &= \sum_{k \in Q} \alpha_k \label{E:N-AC}
\end{align}
where $P$ is the set of edges in the subtree spanned by 1 and 3 and $Q$ is the subtree spanned by all other leaves.  Note that $Q$ must contain all edges in the tree except for the leaf edges associated with 1 and 3.  Thus from
(\ref{E:total}) and (\ref{E:N-AC}) we get
\begin{equation}\label{E:ACleq}
x_1 + x_3 \leq \alpha_1 + \alpha_3.
\end{equation}
Then from (\ref{E:ACgeq}) and (\ref{E:ACleq}) we must have
$$\sum_{k \in P} \alpha_{k} \leq x_1 + x_3 \leq \alpha_1 + \alpha_3.$$
However this cannot be satisfied (the core is empty) unless all of
the internal edge weights in $P$ are zero.  But that every internal edge is in a subtree spanned by pairs of cherries, hence all internal edges weights are zero.  In the latter case, the tree is degenerate and the core is the single element $\vec{\ell}$.
\end{proof}

Notice that for $n=3$, ${\mathcal{T}}$ is always degenerate, and
thus the core will never be empty.

Because the core of tree games is empty in most interesting cases,
the Shapley value is a far more valuable solution concept to
consider.

\ignore{************
 However, the core has the potential to find (or
rule out) degenerate trees easily, unlike the Shapley value.

Suppose we are given the pairwise distances for $n$ leaves of a tree.
If any four leaf subset has an empty core, then the tree is definitely
not degenerate.  But if any of the inequalities hold then the subtree
spanned by the four leaves in the subset contains a degeneracy.  To
illustrate this point, see example \ref{Ex:degen}.

\begin{ex}\label{Ex:degen}

  Consider the 5-leaf tree given in figure \ref{Fi:four&five}.  Let
  $\mu >0$ and $\rho = 0$.  Then the four-leaf subtree ACDE has a
  nonempty core, namely $\left( \begin{array}{c} \alpha+ \mu \\ 0 \\
      \gamma \\ \delta \\ \epsilon \end{array} \right)$.  Thus there
  is a degeneracy among the leaves ACDE which we can see (C, D, E all
  have a common parent).  However, in the four leaf subtree ABCE, we
  have
$$\alpha + \mu + \gamma \leq x_A + x_C \leq \alpha + \gamma$$
so the core is empty.  Thus the tree is not totally degenerate but it
contains a degenerate subtree CDE.

\end{ex}
******************************}


\section{Conclusion}

In this paper we have presented a biological interpretation of the
Shapley value on games derived from phylogenetic trees.  We have
determined the linear transformation $\mathbf{M}$ that produces the
Shapley value from the edge weights of the tree.  We also determined
its null space. It is worth noting again the dependence of these
results on the {\em split counts} of the tree. 
Finally, we characterized the Shapley value
on the space of tree games by four axioms, in much the same way as
Shapley did for the space of all games.

We close the paper with some speculation.  One of our primary
motivations for studying properties of the Shapley value of
phylogenetic tree games was for the possibility of using
game-theoretic concepts to reconstruct trees from data. Our results
on the properties of the Shapley transformation suggest several
directions for further research.  For instance:

\begin{itemize}
\item If there were a way to estimate the Shapley value from data
(such as by quantifying the notion of diversity of populations),
this would be enough to determine edge weights of a degenerate tree.
Do the leaf weights of this tree have any significance?

\item
Is there a way to determine or estimate split counts from data, and
can this assist in determining the correct tree topology? 

\item
Does the converse of Theorem \ref{Thm:nonisomorphic} hold, i.e., if two trees have permutation-equivalent Shapley transformation matrices or permutation-equivalent null spaces, are they isomorphic?

\item
For a given $n$-leaf tree topology, the Shapley transformation takes
a vector of leaf weights to a vector of Shapley values.  However,
one may speak of the space of all weighted $n$-leaf trees (of
various tree topologies), as in \cite{BHVo01}, and we can therefore
view the Shapley transformation as a map (the {\em Shapley map})
from the space of trees to a vector of Shapley values.  However, the
space of trees is naturally embedded in ${\mathbb R}^{{n \choose
2}}$, the space of pairwise distances.  Is there a ``natural''
extension of the Shapley map to this space?  How does the kernel of
the Shapley map extend the null spaces of Theorem \ref{Thm:null}?
Can this map be used to reconstruct trees?

\ignore{******************
 Theorem \ref{Thm:null}
suggests that once a basis for the space of edge weights is fixed,
one may determine the topology of the tree from the null space
$Null(\mathbf{M})$ of its Shapley transformation $\mathbf{M}$,
because every basis element corresponds to a {\em split} of the tree
(i.e., a bi-partition of the leaves) and the topology of the tree
can be uniquely reconstructed from its splits (see \cite{Bune71}).
It is unclear how to choose the ***********************}

\item If we use the Shapley value to rank the species in the Noah's
ark problem for preservation, to what extent can we guarantee that
the diversity of the top $k$ species (i.e., the weight of the
subtree spanning them) approximates the total diversity of all $n$
species?  Determine a bound that depends on $k$ and $n$.
\end{itemize}


\subsection*{Acknowledgements}
The authors thank Susan Holmes and Bernd Sturmfels for helpful feedback regarding
these ideas, and Kyle Kinneberg, Aaron Mazel-Gee, and 
an anonymous referee for valuable comments on an earlier draft.

\bibliographystyle{plain}   
\bibliography{phylo}

\begin{thebibliography}{10}

\bibitem{BHVo01}
Louis~J. Billera, Susan~P. Holmes, and Karen Vogtmann.
\newblock Geometry of the space of phylogenetic trees.
\newblock {\em Advances in Applied Mathematics}, 27(4):733 -- 767, 2001.

\bibitem{DaMc03}
William~H.E. Day and F.R. McMorris.
\newblock {\em Axiomatic Consensus Theory in Group Choice and Biomathematics}.
\newblock SIAM, Philadelphia, 2003.

\bibitem{Fait92}
Daniel~P. Faith.
\newblock Conservation evaluation and phylogenetic diversity.
\newblock {\em Biological Conservation}, 61:1--10, 1992.

\bibitem{Fels04}
Joseph Felsenstein.
\newblock {\em Inferring Phylogenies}.
\newblock Sinauer Associates, Inc., Massachusetts, 2004.

\bibitem{HaSt06}
Klaas Hartmann and Mike Steel.
\newblock Maximizing phylogenetic diversity in biodiversity conservation:
  Greedy solutions to the noah's ark problem.
\newblock {\em Systematic Biology}, 55:644--651, 2006.

\bibitem{Kar02}
Anirban Kar.
\newblock Axiomatization of the shapley value on minimum cost spanning tree
  games.
\newblock {\em Games and Economic Behavior}, 38:265--277, 2002.

\bibitem{MKvH06}
Bui~Quang Minh, Steffen Klaere, and Arndt von Haeseler.
\newblock Phylogenetic diversity within seconds.
\newblock {\em Systematic Biology}, 55:769--773, 2006.

\bibitem{MHCh05}
Arne~\O. Mooers, Stephen~B. Heard, and Eva Chrostowski.
\newblock Evolutionary heritage as a measure for conservation.
\newblock In A.~Purvis, T.~Brooks, and J.~Gittleman, editors, {\em Phylogeny
  and conservation}, pages 120--138. Cambridge Univ.~Press, Cambridgs, UK,
  2005.

\bibitem{myer77}
Roger~B. Myerson.
\newblock Graphs and cooperation in games.
\newblock {\em Mathematics of Operations Research}, 2(3):225--229, 1977.

\bibitem{NePu02}
Klaus Nehring and Clemens Puppe.
\newblock A theory of diversity.
\newblock {\em Econometrica}, 70(3):1155--1198, 2002.

\bibitem{owen86}
Guillermo Owen.
\newblock Values of graph-restricted games.
\newblock {\em SIAM Journal of Algebra and Discrete Mathematics},
  7(2):210--220, 1986.

\bibitem{PaGo05}
Fabio Pardi and Nick Goldman.
\newblock Species choice for comparative genomics: being greedy works.
\newblock {\em PLoS Genetics}, 1(e71):672--675, 2005.

\bibitem{PSOD05}
Sandrine Pavoine, Sébastien Ollier, and Anne-Béatrice Dufour.
\newblock Is the originality of a species measurable?
\newblock {\em Ecology Letters}, 8:579--586, 2005.

\bibitem{SeSt03}
Charles Semple and Mike Steel.
\newblock {\em Phylogenetics}.
\newblock Oxford Univeristy Press, New York, 2003.

\bibitem{Shap53}
Lloyd~S. Shapley.
\newblock A value for n-person games.
\newblock In {\em Ann.~Math.~Studies}, volume~28, pages 307--317. Princeton
  University Press, Princeton, N.J., 1953.

\bibitem{Stee05}
Mike Steel.
\newblock Phylogenetic diversity and the greedy algorithm.
\newblock {\em Systematic Biology}, 54:527--529, 2005.

\bibitem{Weit92}
Martin~L. Weitzman.
\newblock On diversity.
\newblock {\em Quarterly Journal of Economics}, 107(2):363--405, 1992.

\bibitem{Weit98}
Martin~L. Weitzman.
\newblock The {N}oah's ark problem.
\newblock {\em Econometrica}, 66(6):1279--1298, 1998.

\end{thebibliography}

\end{document}